# Three-Stage Adjusted Regression Forecasting (TSARF) for Software Defect Prediction

Shadow Pritchard, MS, The Boeing Company, MO, USA

Bhaskar Mitra, PhD, Pacific Northwest National Laboratory, WA, USA

Vidhyashree Nagaraju, PhD, Stonehill College, MA, USA

*SUMMARY & CONCLUSIONS*

Software reliability growth models (SRGM) [1] enable failure data collected during testing. Specifically, non-homogeneous Poisson process (NHPP) [2] SRGM are the most commonly employed models. While software reliability growth models are important, efficient modeling of complex software systems increases the complexity of models [3]. Increased model complexity presents a challenge in identifying robust and computationally efficient algorithms to identify model parameters and reduces the generalizability of the models.

Existing studies on traditional software reliability growth models [4] suggest that NHPP models characterize defect data as a smooth continuous curve and fail to capture changes in the defect discovery process. Therefore, the model fits well under ideal conditions, but it is not adaptable and will only fit appropriately shaped data. Neural networks and other machine learning methods have been applied to greater effect [5], however limited due to lack of large samples of defect data especially at earlier stages of testing.

In this paper, a *three-stage adjusted regression forecasting model* is proposed to forecast the local regression model [6]. This is a growth curve approximation model that predicts the parameters for a future linear model based on a sliding window of previous linear models. The three stages of the model are as follows:
- Initial fit: Train regression models on a sliding window of the date and record model coefficients.
- Prediction: Fit new regression models to the coefficient lists and predict the value of the next coefficient.
- Error correction: Correct coefficient prediction error using the residual of the last point of the coefficient list and moving average.

The resulting model from the multi-stage process is a forecast of the local regression model that represents the future window of data and is referred to as the predicted line. Results suggest the three-stage model demonstrates better prediction capability compared to existing solutions.

## 1 INTRODUCTION

Non-homogeneous Poisson process (NHPP) [2] software reliability growth models (SRGM) [1] are a common methodology that enables quantitative assessment of software systems by characterizing defect data collected during testing. Traditional NHPP models are developed based on the assumption that the defect detection process can be accurately characterized by a continuous smooth curve. This assumption restricts the model's ability to characterize and predict changes in the failure rate, thus reducing the accuracy of defect prediction. Changepoint models [7] were developed to characterize the changes in the data. However, such models are limited to characterization of the number of changepoints incorporated in the parametric form. Therefore, the mean value function of traditional NHPP SRGM can only exhibit curves based on parametric forms and therefore cannot predict defects accurately beyond a certain level of precision.

Previous studies have proposed several models and methods to improve the predictive accuracy of NHPP models. Dohi et al. developed new models such as the mixed type [8], [9], Markov modulated process [10], Hyper-Erlang [11], logistic regression [12], Poisson regression [13], Cumulative Binomial Trial Process [14] as well as additive [15] and covariate [16] models. Other studies focused on the development of software reliability models with changepoints to characterize changes and stages in a software testing process. Single changepoint models have been developed for hazard rate [4], [17] and failure count data [18], [19]. Most studies present changepoint models for various classes of parametric models such as discrete [20], imperfect debugging [21], fault correction [22], environmental factors [23], [24], test effort [25], and execution time [26] models.

Successive generations of software reliability models possess an increased number of parameters, which requires stable and efficient algorithms for model fitting. Statistical algorithms include the expectation-maximization (EM) [27], expectation conditional maximization (ECM) [28], and hybrid algorithms [3], which combine the EM and ECM algorithms with the Newton-Raphson [29] method. Several studies have also explored the application of soft computing techniques [30], including the genetic algorithm [31] and particle swarm optimization [32], [33]. More recent studies utilize machine learning algorithms [34], specifically variations of neural networks. Promising results based on the artificial neural network (ANN) led to further studies, including evolutionary neural network [35] and ANN with changepoints [36]. Most recently, focus has shifted towards recurrent neural networks (RNN) [37] for software failure prediction [38] along with several variations, including the combined recurrent ANN model [39] for fault detection and correction, dynamic weighted model [40], and Bayesian regularization [41].

This paper proposes a three-stage adjusted regression forecasting (TSARF) model to forecast the local regression model [6]. This is a growth curve approximation model that predicts the parameters for a future linear model based on a sliding window of previous linear models. The three stages of the model are:

- Initial fit: train regression models on a sliding window of the date and record model coefficients.
- Prediction: fit new regression models to the coefficient lists and predict the value of the next coefficient.
- Error correction: correct coefficient prediction error using the residual of the last point of the coefficient list and moving average.

The resulting model from the multi-stage process is a forecast of the local regression model that represents the future window of data and is referred to as the predicted line. Traditional NHPP models, including the Goel-Okumoto (GO) [4], delayed S-shaped (DSS) [41], and Weibull [42] models are considered in order to compare the predictive capability of the TSARF models. All models considered are applied in the context of historical software failure data [43].

The remainder of this paper is organized as follows: Section 2 reviews regression fundamentals, Section 3 details the model. Section 4 lists several measures to assess the goodness-of-fit of model. Section 5 provides illustrations and Section 6 offers conclusion and future research.

## 2 REVIEW OF MODELS

This section provides an overview of regression model, which provides a foundation for the proposed model. A review of software reliability growth models including Goel-Okumoto, Weibull, and delayed S-shaped (DSS) are also provided.

### 2.1 Regression Model

Method of Ordinary Least Squares (OLS) [44] is used to fit the linear regression models to the data.
$$\hat{Y} = X\hat{\beta} \quad (1)$$
where $\hat{Y}$ is the $n \times 1$ matrix of observed response values, $X$ is the $n \times m$ matrix of observed values, and $\hat{\beta}$ is the $m \times 1$ matrix of unknown coefficients. Here $n$ is the number of observations and $m$ is the number of parameters.

The optimal coefficients can be found by minimizing the sum of squared errors:
$$\hat{\beta} = \mathop{\arg\min}_{\beta \in R} \sum_{i=1}^{n}\left(y_i - \sum_{j=1}^{m} x_{i,j}\beta_j\right) \quad (2)$$

where $y_i$ is the $i^{th}$ observation, $x_{i,j}$ is the $i^{th}$ row and $j^{th}$ column of $X$ and $\beta_j$ is the $j^{th}$ coefficient.

Solving for $\beta$ in Equation (2) produces a closed form solution for the optimal $\hat{\beta}$
$$\hat{\beta} = (X^T X)^{-1} X^T \hat{Y} \quad (3)$$

### 2.2 NHPP SRGM

The non-homogeneous Poisson process is a stochastic process [1] that counts the number of events observed as a function of time. In the context of software reliability, the NHPP counts the number of unique faults detected by time $t$. This counting process is characterized by a mean value function (MVF) given by,
$$m(t) = a \times F(t) \quad (4)$$
where $a$ denotes the expected number of unique faults that would be discovered with indefinite testing and $F(t)$ is the cumulative distribution function (CDF) of a continuous probability distribution characterizing the software fault detection process.

#### 2.2.1 Goel-Okumoto (GO) SRGM

The Goel-Okumoto model was originally proposed by Goel and Okumoto [3]. The MVF is
$$m(t) = a(1 - e^{-bt}) \quad (5)$$
where $b$ is the fault detection rate.

#### 2.2.2 Delayed S-shaped (DSS) SRGM

The MVF of the DSS SRGM is
$$m(t) = a(1 - (1 + bt)e^{-bt}) \quad (6)$$
where the term $bte^{-bt}$ can characterize a delay in fault detection induced by phenomenon such as delayed failure reporting and fault masking.

#### 2.2.3 Weibull SRGM

The MVF of the Weibull model is
$$m(t) = a\left(1 - e^{-bt^c}\right) \quad (7)$$
Here $b$ and $c$ are the scale and shape parameters respectively. Setting $c = 1$ in Equation (7) simplifies to the Goel-Okumoto model.

## 3 THREE-STAGE ADJUSTED REGRESSION FORECASTING MODEL

In this section, the model and the stages thereof are discussed.

### 3.1 Initial Fit

In this stage, data is divided, and models fit to each window of data. Then the minimization equation in Equation (2) is modified to
$$\hat{\beta}_w = \mathop{\arg\min}_{\beta \in R} \sum_{i=w}^{w \times k}\left(y_i - \sum_{j=1}^{m} x_{i,j}\beta_j\right) \quad (8)$$
where $\hat{\beta}_w$ is the $w^{th}$ row of the coefficient matrix, $n$ is the length of the data, $k$ is the window size, and $w$ is the number of the window. All the coefficients are recorded in a $\frac{n}{k} \times m$ matrix to be used in the next stage.

## 3.2 Prediction

The prediction stage uses the previously found $\hat{\beta}$ values to forecast what the coefficients should be for the future window. Each coefficient must be predicted so for the $\rho^{th}$ coefficient the minimization equation is

$$\hat{B}_\rho = \underset{B \in R}{\arg min} \sum_{i=1}^{n/k} \left( \hat{\beta}_{i,\rho} - \sum_{j=1}^{m} i \times B_j \right) \quad (9)$$

where $\hat{B}$ is the coefficients for the model to forecast the coefficients for the time series. Here $i$ is used in the place of $x$ as the goal is to predict the $\left(\frac{n}{k}+1\right)^{th}$ coefficient for the predicted line.

Plugging $\hat{B}$ into the regression equation, the $\left(\frac{n}{k}+1\right)^{th}$ coefficients can be predicted:

$$y_\rho = \hat{\beta}_{\rho,0} + \left(\frac{n}{k}+1\right) \times \hat{\beta}_{\rho,1} \quad (10)$$

where $y_\rho$ is the forecasted value for the $\rho^{th}$ coefficient for the predicted line.

## 3.3 Error Correction

The initial value of $y_\rho$ is typically a miscalculation of the actual value. This results in models that have predicted lines that could have negative slope or intercept that shifts the line away from the actual trend of the data. However, the error can be estimated and in conjunction with previous coefficient values used to correct the error.

$$\epsilon = B_{\frac{n}{k},\rho} - \left( \hat{\beta}_{\rho,0} + \left(\frac{n}{k}\right) \times \hat{\beta}_{\rho,1} \right) \quad (11)$$

where $B_{\frac{n}{k},\rho}$ is the actual fitted value for the last model fit in training and the rest of the term is predicted value for that coefficient. $y_\rho$ is then updated by adding this term

$$y_\rho = y_\rho + \epsilon \quad (12)$$

Finally, a moving average (MA) of the coefficients is used to further adjust the predicted coefficient values. The number of previous values to use in the moving average can be input manually or determined algorithmically. The process to determine the number of moving averages looks at the last window of the training data and finds the moving average with the least mean squared error.

After determining the moving average, it is applied, and the predicted value is giving extra weight compared to the MA correction.

$$y_\rho = 0.5 \times \left( y_\rho + \frac{1}{d} \sum_{i=1}^{d} B_{\frac{n}{k}-i} \right) \quad (13)$$

where $d$ is the number of previous values to use in the moving average.

## 4 GOODNESS-OF-FIT MEASURES

This section summarizes goodness of fit measures to assist in model comparison based on predictive capability.

### 4.1 Predictive Mean Squared Error (PMSE)

Predictive mean squared error is same as mean squared error (MSE) but only records result over the testing set of data of length $n - k$. PMSE is a much more useful metric than MSE as the model is intended to forecast more so than interpolate.

$$PMSE = \frac{1}{n-k} \sum_{i=k+1}^{n} (f(x_i) - y_i)^2 \quad (14)$$

where $f(x_i)$ is the predicted value and $y_i$ is the actual value.

### 4.2 Predictive Ratio Risk (PRR)

The predictive ratio risk of a model is

$$PRR = \sum_{i=k+1}^{n} \left( \frac{f(x_i) - y_i}{f(x_i)} \right)^2 \quad (15)$$

where the term in the denominator penalizes underestimation of the number of defects more heavily than an overestimate.

### 4.3 Predictive Power (PP)

The predictive power of a model is

$$PP = \sum_{i=k+1}^{n} \left( \frac{f(x_i) - y_i}{y_i} \right)^2 \quad (16)$$

Where the term in the denominator penalizes overestimation of the number of defects.

## 5 ILLUSTRATIONS

In this section, the results of the model are recorded on two datasets, CSR3 and S2 roughly following exponential and S-shape respectively. Datasets are taken from the Handbook of Software Reliability Engineering [43]. CSR3 and S2 have 104 and 54 failures recorded. Each data set is a list of the times software failures were discovered and the list is converted into a growth curve of the cumulative number of failures discovered at a given time. Initial values for the size of the sliding windows $k$ are set to 10% of the size of the training data set so that 10 windows of data are used and the $11^{th}$ is predicted.

### 5.1 Model Fit and Goodness-of-fit Measures

Figure 1 shows the three-stage adjusted regression forecasting model fit to CSR3. Blue lines are the initial fit models and red line is the predicted.
As can be seen, the fit is close to the actual points and strongly resembles what the actual linear regression line would be.

Table 1 displays the statistical measures of the TSRAF model compared to GO, Weibull, and DSS models.

*Table 1 – CSR3 Results*

| Model | PMSE | PRR | PP |
|---|---|---|---|
| TSARF | **0.85** | **0.001** | **0.001** |
| DSS | 32.657 | 0.044 | 0.038 |
| GO | 15.804 | 0.020 | 0.018 |
| Weibull | 6.951 | 0.009 | 0.008 |

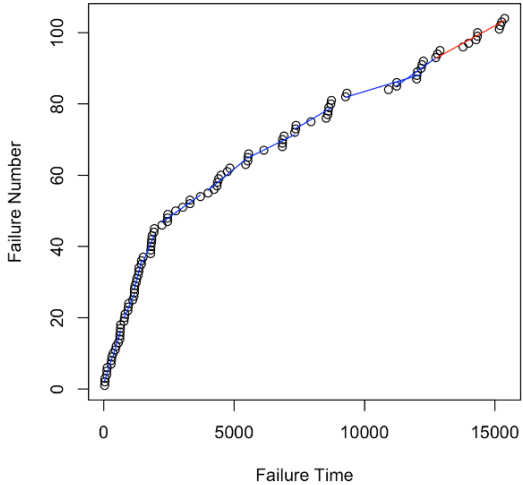

Figure 1: Model Fit

TSARF significantly outperforms previous statistical models that best fit CSR3 data. The data has an appearance rather like linear data with a changepoint at 2,500 time units, which could contribute to the performance of this algorithm and poor performance of single curve fitting methods.

To test the ability of the algorithm to adapt to different distributions, Table 2 shows statistical measures of the TSRAF model compared to GO, Weibull, and DSS models when applied to S2 data, which is s-shaped.

*Table 2 – S2 Results*

| Model | PMSE | PRR | PP |
|---|---|---|---|
| TSARF | 5.91 | 0.015 | 0.016 |
| DSS | 9.583 | 0.028 | 0.024 |
| GO | **2.667** | **0.007** | **0.007** |
| Weibull | 4.425 | 0.012 | 0.011 |

Unlike exponential data, TSARF does not predict as well for the s-shaped data. This, however, is due to the parameters used and the method by which they are selected. A refinement of the parameter selection can greatly decrease the error as shown in the next section.

5.2 *Parameter Sensitivity*

In this section the model parameters are changed to evaluate how that changes the results. The parameter values are typically selected automatically by algorithmic processes, but a sensitivity analysis is done to further understand how changing the values influences the outcome.

5.2.1 *Window Size*

Table 3 shows the changes to PMSE on CSR3 and S2 when the window size of TSARF is changed.

*Table 3 – Window Size PMSE Sensitivity*

| Size | CSR3 | S2 |
|---|---|---|
| 4 | 45.65 | 4.80 |
| 5 | 13.83 | 13.85 |
| 6 | 6.31 | 31.66 |
| 7 | 50.44 | 18.91 |
| 8 | 14.16 | 14.07 |
| 9 | 0.85 | 109.68 |
| 10 | 2.80 | 7.78 |
| 11 | 13.35 | 5.86 |
| 12 | 22.30 | 3.66 |

As listed in Table 3, the ideal window size for CSR3 is 9, but 4 or 12 is much better for S2. Incidentally, the floor of 10% of the length of each training set is 9 and 4 respectively.

5.2.2 *Moving Average*

Table 4 shows variation in PMSE on CSR3 and S2 when the length of the moving average of TSARF is changed.

*Table 4 – Moving Average PMSE Sensitivity*

| Length | CSR3 | S2 |
|---|---|---|
| 1 | 1103.91 | 134.13 |
| 2 | 110.15 | 68.55 |
| 3 | 4.23 | 4.80 |
| 4 | 0.85 | 22.70 |
| 5 | 3.52 | 540.32 |
| 6 | 104.69 | 1725.57 |

The difference for changing the moving average length is much more drastic than the changes to window size. CSR3 performs quite poorly until a moving average of 3, 4, or 5 is used, and then the PMSE is low. S2 is similar, but with a smaller window for good moving average length of 3 or 4.

## 6 CONCLUSION AND FUTURE RESEARCH

This paper proposed a three-stage adjusted regression forecasting model for prediction of software defects with high accuracy using a sliding window approach. The proposed model was compared with three popular traditional software reliability growth models including Goel-Okumoto, Weibull, and delayed S-shaped based on several predictive measures. Results on two datasets CSR3 and S2 suggest that the proposed model demonstrates better predictive ability than traditional models. However, optimal window size and moving average length may not be found with the proposed methods. Thus, poor fitting may be corrected through user identified parameters.

Future work includes a refinement of the algorithm and the parameter selection methods.

## BIOGRAPHIES

Shadow Pritchard, MS
Software Engineer
The Boeing Company
4220 Duncan Ave, St. Louis, MO 63110

e-mail: swp7196@utulsa.com

Shadow Pritchard is a software engineer at Boeing. He received his MS (2022) in Computer Science at The University of Tulsa., where he was a recipient of the Team 8 Cyber Fellows scholarship. His interests and expertise include machine learning, software reliability, artificial intelligence, and statistical modeling.

Bhaskar Mitra, PhD
Power Systems Engineer
Pacific Northwest National Laboratory
902 Battelle Blvd, Richland, WA 99354

e-mail: bhaskar.mitra@pnnl.com

Dr. Bhaskar Mitra (IEEE SM '23) is a Power Systems Engineer with the Distribution Systems group at Pacific Northwest National Laboratory (PNNL). He earned his master's and doctorate in electrical engineering at University of No at Charlotte and his bachelor's, also in electrical engineering, at West Bengal University of Technology. He has the experience of working with several Department of Energy Offices namely, OE, WTO, VTO and SETO. His research includes development of fault detection algorithms using advanced signal processing techniques, load modelling, and development of optimization framework for resource integration. Currently he is the chapter lead for CIGRE B4/A3.8 Group.

Vidhyashree Nagaraju, PhD
Assistant Professor
Department of Computer Science
Stonehill College
320 Washington St, North Easton, MA 02357

e-mail: vnagaraju@stonehill.edu

Vidhyashree Nagaraju is an Assistant Professor in the Department of Computer Science at the Stonehill College. She received her PhD (2020) in Computer Engineering from the University of Massachusetts Dartmouth.